\documentclass[prl,superscriptaddress,twocolumn,10pt,nofootinbib,aps, floatfix]{revtex4-1}
\usepackage{amssymb}
\usepackage{amsmath}
\usepackage{appendix}
\usepackage{times}
\usepackage{graphicx}
\usepackage{subeqnarray}
\newcommand{\ket}[1]{|#1\rangle}
\newcommand{\bra}[1]{\langle #1|}
\usepackage{color}

\begin{document}

\title {Doubled Quantum Spin Hall Effect with High-Spin Chern Number in $\alpha$-Antimonene and $\alpha$-Bismuthene}

\author{Yingxi Bai}
\author{Linke Cai}
\author{Ning Mao}
\author{Runhan Li}
\author{Ying Dai}
\email{daiy60@sdu.edu.cn}
\author{Baibiao Huang}
\author{Chengwang Niu}
\email{c.niu@sdu.edu.cn}
\affiliation
{School of Physics, State Key Laboratory of Crystal Materials, Shandong University, Jinan 250100, China}

\begin{abstract}
The discovery of quantum spin Hall effect has ignited the field of topological physics with vast variety of exotic properties. Here, we present the emergence of doubled quantum spin Hall effect in two dimensions characterized with a high spin Chern number of ${\mathcal C_S}=2$ and two pairs of helical edge states. Although is overlooked and invisible in topological quantum chemistry and symmetry indicator theory, the already experimentally synthesized $\alpha$-antimonene and $\alpha$-bismuthene are revealed as realistic material candidates of predicted topological states with band inversions emerging at generic $k$-points, rather than the high-symmetry momenta. Remarkably, the nontrivial energy gap can be as large as 464 meV for $\alpha$-bismuthene, indicating the high possibility of room-temperature observation of the doubled quantum spin Hall effect. Moreover, a four-band effective model is constructed to demonstrate further the feasibility of attaining this type of nontrivial topology. Our results not only uncover a novel topological character of antimony and bismuth, but will also facilitate the experimental characterization of the previously overlooked hidden topology.
\end{abstract}

\maketitle

Topological quantum states as the focus of recent interests have reshaped our understanding of physics and materials, with promising applications in emerging topotronics~\cite{Hasan2010,Qi2011,Bansil2016rmp,Xiao2021NRP,Breunig2021}. The conceptual milestone is the $\mathbb{Z}_2$ topological insulators (TIs) distinguished by an nonzero $\mathbb{Z}_2$ invariant and odd pairs of helical edge/surface states arising from the time-reversal symmetry $\mathcal T$~\cite{Hasan2010,Qi2011,Bansil2016rmp,Xiao2021NRP,Breunig2021}. In fact, since the birth of $\mathbb{Z}_2$ TIs in two dimensions, also known as the quantum spin Hall (QSH) insulators~\cite{Kane,Bernevig2006}, a variety of topological quantum states have been constantly proposed and intensively explored, from the topological crystalline insulators to topological semimetals even with different bulk-boundary correspondences, and remarkably to the antiferromagnetic TIs~\cite{Fu2011prl,Hsieh,Liu2014NM,Niu2015NL,Wan2011,Wang2012,Weng2015,Huang2015,Yu2015prl,Kim2015,Schindler2018,ZhangDPRL2019,Li2019SA,Niu2020prl}. Generally, band inversion, a heuristic scenario to understand the nontrivial bulk topology, occurs at the high-symmetry points, and thus the wavefunction information at the high-symmetry points provides a convenient way to reveal the bulk topology in topological quantum chemistry and symmetry indicator theories~\cite{Bradlyn2017,Po2017,Song2018,Watanabe2018,Tang2019}. However, physically, the symmetry indicators may provide an incomplete topological characterization, for example, the Weyl points at generic $k$-points without the inversion or $S_4$ symmetry usually cannot be described ~\cite{HughesPRB2011,TurnerPRB2012,GAO2021667}, and unexpected hidden topology is still possible even if the symmetry indicator is trivial. Although considerable efforts have been devoted to, such hidden topology with stable nontrivial phase has yet been proposed~\cite{Po2018PRL,Radha2021PRB,Khalaf2021PRR}.

As a wonderful platform to investigate the emergent topological phenomena, bismuth- and antimony-based materials in both the two and three dimensions have recently drawn significant attentions, such as the TIs in Bi$_{1-x}$Sb$_x$~\cite{Hsieh2008Nature}, Sb/Bi chalcogenides~\cite{Zhang2009np,Xia2009NP}, Tl(Sb/Bi)Se$_2$~\cite{HsinPRL2010,SatoPRL2010,KurodaPRl2010}, Bi$_4$(Br/I)$_4$~\cite{zhou2014,Autes2016}, and Bi(111) bilayer~\cite{LiuZPRL2011,zhoum,Niu2015,Reis2017}, dual TI in Bi$_1$Te$_1$~\cite{eschbach2017bi}, Dirac Semimetal in Na$_3$Bi~\cite{Wang2012,liu2014discovery}, antiferromagnetic TIs in MnBi$_2$Te$_4$~\cite{ZhangDPRL2019,Li2019SA}. Moreover, even though has been long thought to be topologically trivial, recent works uncover that the bulk Bi is in fact both a higher-order TI~\cite{Schindler2018} and a first-order topological crystalline insulator~\cite{Hsu2019PNAS}. In two dimensions, Bi(111) bilayer is unstable with respect to a experimentally observed structural transformation into the buckled black-phosphorous structure~\cite{Nagao2004PRL}, i.e., $\alpha$-bismuthene ($\alpha$-Bi), but which becomes topologically trivial as previously reported~\cite{Wada2011PRB}. Meanwhile, the $\alpha$-antimonene ($\alpha$-Sb) has also been intensively explored without nontrivial topology~\cite{Zhang2016ac}.

In the present work, we uncover a long-awaited hidden topology of doubled QSH effect in $\alpha$-Sb and $\alpha$-Bi. Unlike the band inversions in regular topological insulators and/or topological semimetals at high-symmetry points, the band inversions induced by spin-orbit coupling (SOC) in $\alpha$-Sb and $\alpha$-Bi emerge at generic $k$-points. Remarkably, based on first-principles calculations and effective model analysis, we show two pairs of gapless helical edge states with the doubled QSH effect can be characterized by a high spin-Chern number of ${\mathcal C_S}=2$ and gapless Wannier charge centers (WCC). The nontrivial energy gap are 91 meV and 464 meV for $\alpha$-Sb and $\alpha$-Bi, respectively, which are well above the energy scale for room temperature applications. Our work provides a realistic exception to the topological quantum chemistry and symmetry indicator theory with experimentally feasible examples.
 
The density functional calculations are performed within the generalized gradient approximation (GGA) of Perdew-Burke-Ernzerhof (PBE)~\cite{Perdew} as implemented in the Vienna ab initio simulation package (VASP)~\cite{Kresse} and the FLEUR code~\cite{fleur}. A vacuum layer of 20 \AA~ is used to avoid interactions between nearest slabs for VASP. The kinetic energy cutoff is fixed to 500 eV, and all structures are relaxed until the residual forces are less than 0.01 eV/\AA. The criterion of total energy for convergence is set as $10^{-6}$ eV. Spin-orbit coupling (SOC) is included in the calculations self-consistently. The phonon calculations are carried out by using the density functional perturbation theory as implemented in the PHONOPY package~\cite{TOGO2015SM}. The maximally localized Wannier functions (MLWFs) are constructed using the Wannier90 code~\cite{Mostofi2008}. 

\begin{figure}
\centering
\includegraphics{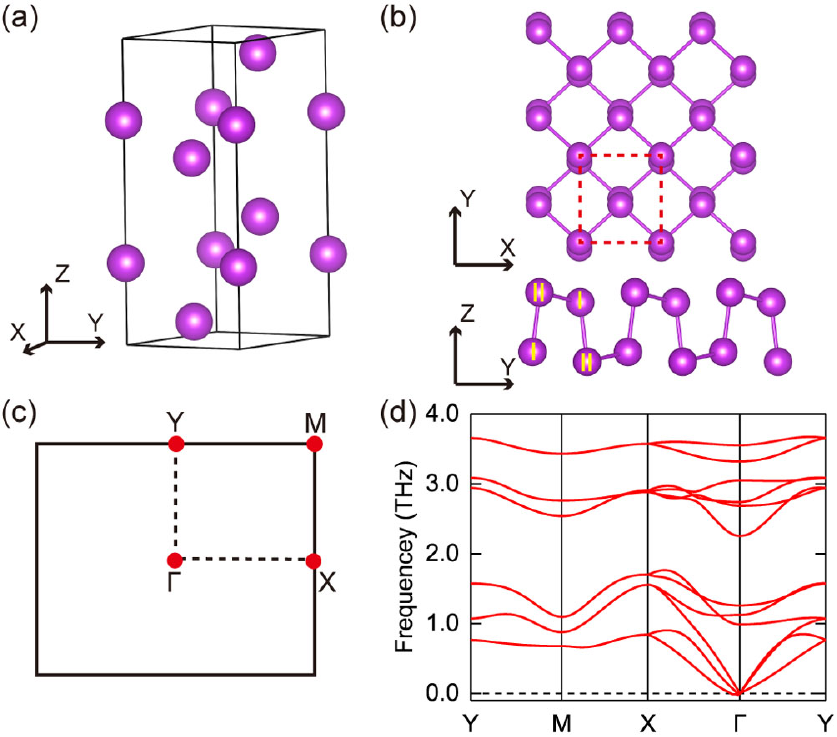}
\caption{(a) Crystal structure of bulk Sb/Bi in space group $R\bar{3}m$. (b) Top and side views of $\alpha$-Sb/Bi in buckled black phosphorene structure with the unit cell indicated by red dashed lines. Labels in the side view present different categories of Sb/Bi atoms. (c) The Brillouin zone for the unit cell with marked high symmetry points. (d) Phonon spectra for $\alpha$-Bi, indicating that the $\alpha$-Bi is dynamically stable.}
\label{str}
\end{figure}

Bulk Sb and Bi crystalize in a rhombohedral structure with space group $R\bar{3}m$, as shown in Fig.~\ref{str}(a), and exhibit a layered structure stacked along the $z$ axis. Each layer forms a buckled hexagonal lattice. However, below a certain thickness, the Sb and Bi thin films grow in the black-phosphorous structure, i.e., resembles the pseudocubic (110) surface, namely $\alpha$-antimonene and $\alpha$-bismuthene. As illustrated in Fig.~\ref{str}(b), there are four atoms in a primitive unit cell, and distinctly different from the two flat atomic layers in phosphorene, $\alpha$-Sb and $\alpha$-Bi are buckled due to the reduced sp$^3$ hybridization as compared with P, which have confirmed by the experimental characterizations and been obtained on different substrates~\cite{Nagao2004PRL,Bianchi2012PRB,Gou2020SA}. Our total energy investigations indicate that the buckled structures are energetically more favorable than the unbuckled one by 57.5 meV and 36.3 meV for Sb and Bi, respectively. Interestingly, this buckling breaks the space inversion symmetry, resulting in the first example of elemental ferroelectric materials~\cite{Xiao2018AFM}. The buckled $\alpha$-Sb and $\alpha$-Bi crystalline in the space group $Pmn2_1$, other than the $Pmna$ for the unbuckled phosphorene. Four atoms in each unit cell can be divided into two categories with different Wyckoff positions, labeled as $I$ and $II$ in the bottom panel of Fig.~\ref{str}(b). The optimized lattice constants are $a=4.35~\rm\AA, b=4.73~\rm\AA$ for $\alpha$-Sb and $a=4.56~\rm\AA, b=4.89~\rm\AA$ for $\alpha$-Bi. The phonon spectrum calculations, with all positive phonon branches in the entire Brillouin zone, indicate the $\alpha$-Sb and $\alpha$-Bi are dynamically and difficult to destroy once formed (see Fig.~\ref{str}(d) and Ref.~\cite{supp}). 

\begin{figure}
\centering
\includegraphics{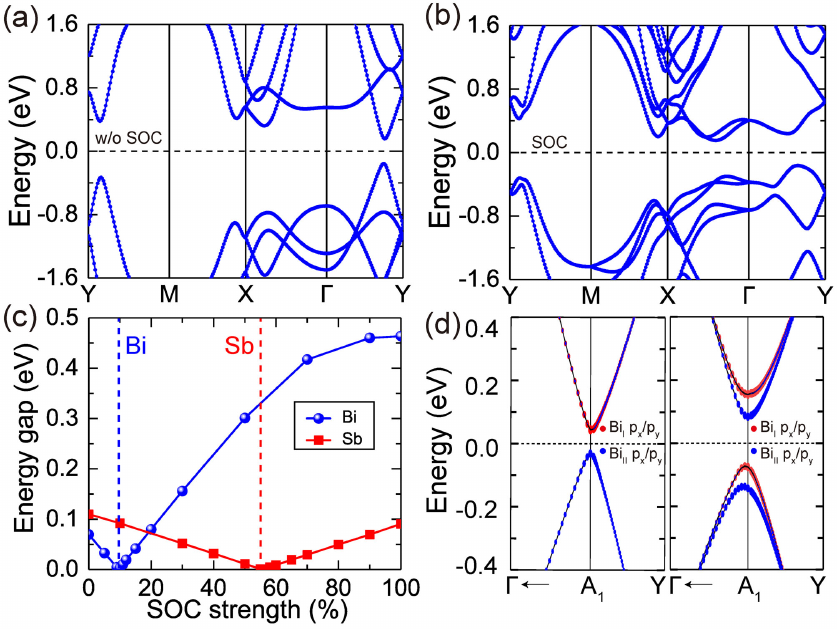}
\caption{Band structures of $\alpha$-Bi (a) without and (b) with SOC along the high symmetry lines. The Fermi level is indicated with a dashed	line. (c) Variation of the energy gaps for $\alpha$-Sb and $\alpha$-Bi as a function of SOC strength. The gap closure emerges with four band-crossings locate at generic $k$-points. (d) Orbitally resolved band structures for $\alpha$-Bi (left panel) without and (right panel) with 30\% SOC around a generic $k$-point A$_1$ (0.030, 0.375).}
\label{band}
\end{figure}

We now focus on the band topology and present the band structures of $\alpha$-Bi without and with SOC in Figs.~\ref{band}(a) and~\ref{band}(b). As we can see, the $\alpha$-Bi is an insulator both without and with SOC. The valence and conduction bands near the Fermi level are mainly contributed by the $p_x$ and $p_y$ orbitals, and there is no band inversion at all of the high-symmetry momenta. However, close examination of the band structures in whole 2D Brillouin zone versus SOC reveal signs of a topological phase transition. Figure~\ref{band}(c) displays the variation of the energy gap as the SOC strength is varied. Starting from a calculation without SOC, it is clearly shown that the global energy gap decreases with increasing SOC strength, and a gap-closing emerges when the SOC strength is 10\% of the normal Bi (55\% for Sb). With further increasing SOC, the energy gap reopens and increases rapidly, reaching as much as 464 meV for $\alpha$-Bi (91 meV for $\alpha$-Sb) under normal SOC strength. Physically, a topological phase transition with band inversion is often revealed by such a gap closing and reopening phenomenon. 

To elucidate the topological phase transition more clearly, Fig.~\ref{band}(d) presents the orbitally resolved band structures without and with 30\% SOC. We note that the valence band maximum (VBM) and conduction band minimum (CBM) locate at a general $k$ point A$_1$, which shifts slightly accompanying the variation of the SOC strength, rather than the high symmetry points and/or lines. In the absence of SOC, the CBM is dominated by the Bi$_{\rm {I}}$-$p_{x/y}$ while the VBM is dominated by the Bi$_{\rm {II}}$-$p_{x/y}$ with a direct energy gap of 71 meV. When a 30\% SOC is switched on, the orbital characters at the A$_1$ point are inverted, but which does not change the symmetry eigenvalues at the high-symmetry points. Therefore, the $\alpha$-Bi is topologically trivial according to the topological quantum chemistry~\cite{supp}, agreement with the previous reports~\cite{Wada2011PRB}. On the other hand, it is apparent that, in this case, the bulk topology can not be indicated by the symmetry eigenvalues.

\begin{figure}
\centering
\includegraphics{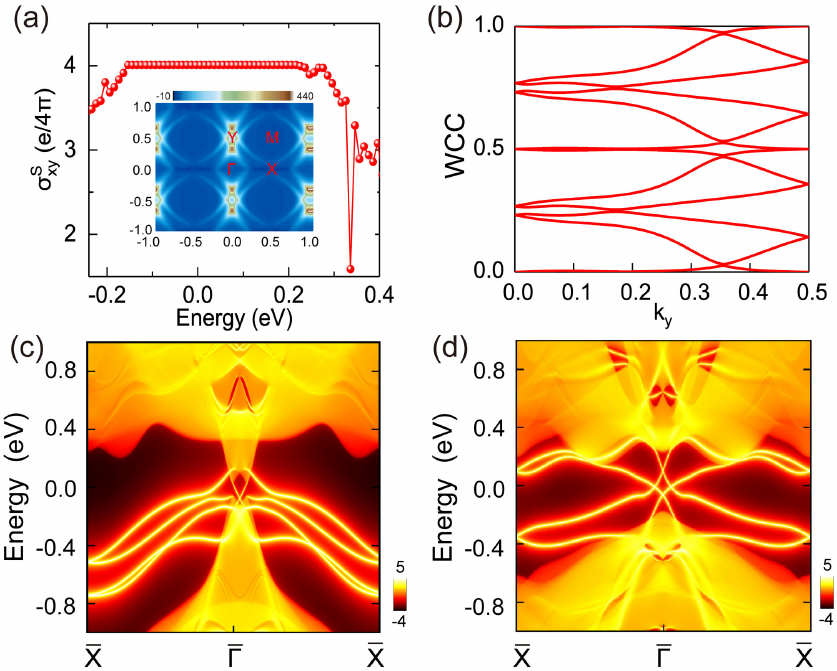}
\caption{(a) Spin Hall conductivity $\sigma_{xy}^{S}$ of $\alpha$-Bi versus the position of the Fermi level E$_F$, showing a quantized value within the energy window of the SOC gap. (Inset) K-space distribution of spin Berry curvature $\Omega^{S}({\bf k})$ within the SOC gap. (b) Evolution of WCC for $\alpha$-Bi. The calculated edge states of (c) $\alpha$-Sb and (d) $\alpha$-Bi with two pairs of gapless edge states.}
\label{edge}
\end{figure}

To identify the topologically nontrivial character with band inversions at generic k-points, we consider the occupied bands in the whole momentum space and calculate both the spin Chern number $\mathcal C_S$ and spin Hall conductivity $\sigma_{xy}^{S}$~\cite{Sinova,Yang2011prl}. In a respective insulating region, $\mathcal C_S$ is analogous to $\sigma_{xy}^{S}$ with the alternative representation $\sigma_{xy}^S=2\mathcal C_S \cdot e/(4\pi)$, and it is well known that $\mathcal C_S =1$ provides an equivalent characterizatio to $\mathbb{Z}_2 =1$ for the QSH insulators~\cite{Yang2011prl}. Although spin is no longer a good quantum number after including SOC, $\mathcal C_S$ can still be defined, as $\mathcal C_S =(\mathcal C_+ - \mathcal C_-)$/2, if the spectrum of the projected spin operator is gaped~\cite{Yang2011prl}, and quite recently has been experimentally detected~\cite{Lv2021PRL}.  $\mathcal{C}_+$ and $\mathcal{C}_-$ are Chern numbers of the spin-up and spin-down manifolds given by ${\mathcal C_{\pm}}=1/(2\pi)\int_{BZ} \Omega_{\pm}({\bf k}) d^{2}k$, where $\Omega_{\pm}({\bf k})$ are the Berry curvatures of all occupied bands constructed from respective spin states, calculated according to $\Omega_{\pm}({\bf k})=2{\rm Im}\sum_{n}^{occ} \langle \partial_{k_x} u_{kn}^{\pm} | \partial_{k_y} u_{kn}^{\pm} \rangle$~\cite{Yao2004}. For which, the MLWFs that reproduce the band dispersion of $\alpha$-Bi precisely are constructed. The calculated Chern numbers of two spin manifolds are ${\mathcal C_+} = 2$ and ${\mathcal C_-} = -2$, resulting in a spin Chern number ${\mathcal C_S}=2$. To clearly visible the quantization, Fig.~\ref{edge}(a) presents the spin Hall conductivity, given by $\sigma_{xy}^{S}=e\hbar\int\frac{d^2 k}{(2\pi)^2}\, \Omega^{S}({\bf k})$, as a function of the Fermi level. The integrand $\Omega^{S}({\bf k})$ is the spin Berry curvature of all occupied bands below the Fermi level
\begin{eqnarray}
\Omega^{S}({\bf k})=-2{\rm Im} \sum_{m\ne n}\frac{\bra{\psi_{m{\bf k}}}J_{x}^{s}\ket{\psi_{n {\bf k}}}\bra{\psi_{n{\bf k}}}\upsilon_{y}\ket{\psi_{m {\bf k}}}}{(E_{n\bf k}-E_{m\bf k})^2} \, ,
\end{eqnarray}
and $J_{x}^{s}=(\hbar/4)\{\sigma^z, \upsilon_x\}$ represents a spin current flowing into the $x$ direction with spin polarization perpendicular to the plane. One can clearly see the quantization of $\sigma^{S}_{xy}=4$, which arises mainly from the $\Omega^{S}({\bf k})$ around the Y points as illustrated in the insert of Fig.~\ref{edge}(a), within the energy window of SOC gap. Just like the higher Chern numbers for quantum anomalous Hall effect~\cite{Wang2013prl}, the ${\mathcal C_S}$ and $\sigma_{xy}^{S}$ are higher and doubled compared to that of the conventional QSH effect, revealing the emergence of doubled QSH effect~\cite{Wang2016nat}, a hidden topology beyond the topological quantum chemistry and symmetry indicator theory. This is further explicitly confirmed by the calculated WCC as presented in Fig.~\ref{edge}(b). Although the $\mathbb{Z}_2$ invariant is equal to 0, one can clearly find that there is no gap in the WCC spectrum.

\begin{figure}
\centering
\includegraphics{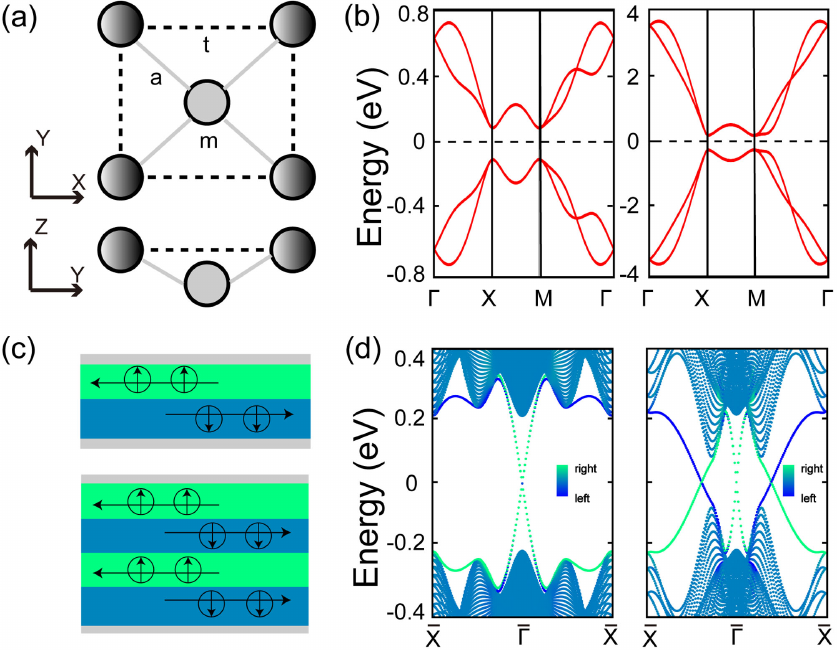}
\caption{(a) Schematic of the four-band model on a 2D square lattice. (b) Insulating band structures and (d) topological edge states for (left panel) conventional quantum spin Hall effect (QSHE) and (right panel) doubled QSHE, which are characterized by one pair and two pairs of helical edge state as sketched in (c). The color transition from blue to green in (d) represents the weight of atoms located	from left to right side of the nanoribbon structure.}
\label{model}
\end{figure}

It is well known that one of the most exotic signatures for the conventional QSH effect with ${\mathcal C_S}=1$ is the existence of one pair of gapless helical edge states~\cite{Hasan2010,Qi2011,Bansil2016rmp}. Characterized by a doubled magnitude of ${\mathcal C_S}=2$, we now show that the doubled QSH effect hosts two pairs of metallic helical edge states, i.e., two right-going edge modes with spin up and two left-going ones with spin down [see Fig.~\ref{model}(c)]. To illustrate this, we carry out calculations of the edge-state band structures using the MLWFs of $\alpha$-Sb and $\alpha$-Bi with their spin polarization checked by computing the expected value of Pauli matrices $\sigma_z$ in the basis of MLWFs. Figures~\ref{edge}(c) and ~\ref{edge}(d) present the edge spectra of semi-infinite $\alpha$-Sb and $\alpha$-Bi along the (110) direction [see Fig.~\ref{str}(b)], respectively. Clearly, one can see that two pairs of nontrivial edge states connect the conduction and valence bands and cross at the $\bar \Gamma$ and $\rm{\bar X}$ points for both of $\alpha$-Sb and $\alpha$-Bi. In addition, as in the case of conventional QSH effect with ${\mathcal C_S}=1$, two pairs of edge states for doubled QSH effect are spin polarized with the directions of spin polarization locked with their momentum, and thus the carriers with opposite spins move in opposite directions on a given edge~\cite{supp}.

Given the ideal experimentally synthesized material candidates, we then proceed to prove that the nontrivial topology of doubled QSHE is indeed possible by a four-band effective model on a 2D square lattice with the Hamiltonian written as
\begin{align}
\begin{split}
H=& -[(m+t (\cos k_{x} +\cos k_{y})] \tau_{z} + a\tau_{x}\beta_k+\\
& \lambda\tau_x[1/2sink_x \sigma_x +  sink_y \sigma_y], \\  
\end{split}
\end{align}
where $\beta_k $ is a polynomial of the momentum $\bf{k}$ with $\beta_k = [1+e^{ik_x} + e^{ik_y}+e^{i(k_x+k_y)}] $ and    $\sigma$/$\tau$ denote for the vectors of spin/sublattice Pauli matrices. Besides, the onsite energy, SOC strength, and magnitudes of the nearest-neighbor and next-nearest-neighbor hopping correspond to parameters $m$, $\lambda$, $a$, and $t$, respectively. In our calculation, all the parameters are scaled with the onsite energy parameter $m$, and $t$ is fixed to be $t=m$.

When the SOC effect is ignored, $\lambda=0$, the model system hosts a Dirac semimetal phase with the Dirac points located along the high-symmetry lines X-M and Y-M. As expected, similar to graphene, a band gap opens when the SOC is turned on as illustrated in the left panel of Fig.~\ref{model}(b) with $\lambda=2.5m$ and $a=1.5m$ as an example. A QSH insulator with ${\mathcal C_S}=1$ ($\mathbb{Z}_2 =1$) is obtained. In fact, this conventional QSH effect can be realized in unbuckled $\alpha$-Bi~\cite{supp}. As shown in the left panel of Fig.~\ref{model}(d), a pair of gapless edge states emerge within the nontrivial gap, in direct agreement with the calculated values of ${\mathcal C_S}=1$. Remarkably, the nontrivial energy gap can be engineered using the nearest-neighbor hopping parameter $a$, and the gap closes at $a = 3.05m$. It should be noted that the band crossing points are located at generic $k$-points, rather than the high-symmetry ones, just like the materials do. When $a$ is further increased, the energy gap reopens and its size increases with $a$, whereby it mediates a topological phase transition from the conventional QSH effect to the double QSHE accompanying the helical edge states changes from one pair to two pairs as sketched in Fig.~\ref{model}(c). The left panel of Fig.~\ref{model}(b) present the band structure with $a=4m$, an insulating gap is clearly shown with its nontrivial topology verified by evaluating the WCC. Moreover, two pairs of gapless edge states, the key manifestation of the doubled QSHE, emerge on both the left and right sides of the nanoribbon that can be clearly distinguished from the projected bulk states as shown in the right panel of Fig.~\ref{model}(d). In addition, to further establish the existence of doubled QSH effect, we propose a tight-binding model by considering the $p_x$ and $p_y$ orbitals based on the MagneticTB package~\cite{Zhang2022cpc}. After including SOC, the electronic band structure reveals that an insulator can indeed be obtained, where the gapless WCC spectrum and two pairs of metallic edge states demonstrate explicitly the topologically nontrivial nature and the doubled QSH effect~\cite{supp}.

In conclusion, we theoretically demonstrated the emergence of doubled QSH insulators in experimentally synthesized $\alpha$-Sb and $\alpha$-Bi, which are overlooked as trivial insulators previously. Similar to the high Chern numbers for quantum anomalous Hall effect, a high spin Chern number of ${\mathcal C_S}=2$ is used to characterize the nontrivial topology. Remarkably, two pairs of helical edge states emerge within the nontrivial gap. Our results greatly enrich the physics and expand the domain of topological phases beyond the topological quantum chemistry and symmetry indicator theory, which are expected to draw immediate experimental attentions.

\acknowledgments{This work was supported by the National Natural Science Foundation of China (Grants No. 11904205, No. 12074217, and No. 12174220), the Shandong Provincial Natural Science Foundation of China (Grants No. ZR2019QA019 and No. ZR2019MEM013), the Shandong Provincial Key Research and Development Program (Major Scientific and Technological Innovation Project) (Grant No. 2019JZZY010302), and the Qilu Young Scholar Program of Shandong University.}

%\bibliographystyle{apsref}
%\bibliography{my_bibliography}


\end{document}